\documentclass[prb,showpacs,twocolumn]{revtex4}

\usepackage[dvips]{graphicx}
\usepackage{amsmath}

\usepackage{color}
\begin{document}

\title{Effect of pressure on the Electron Spin Resonance of a Heavy-Fermion metal}

\author{J.~Sichelschmidt}
\affiliation{Max Planck Institute for Chemical Physics of Solids,
D-01187 Dresden, Germany}

\author{H.-A.~Krug~von~Nidda}

\affiliation{Experimental Physics V, Center for Electronic Correlations and Magnetism,
University of Augsburg, D-86135 Augsburg, Germany}

\author{D.~V.~Zakharov}

\affiliation{Experimental Physics V, Center for Electronic Correlations and Magnetism,
University of Augsburg, D-86135 Augsburg, Germany}

\author{I.~Fazlizhanov}
\affiliation{E.~K.~Zavoisky Physical Technical Institute, 420029
Kazan, Russia}

\author{J.~Wykhoff}
\affiliation{Max Planck Institute for Chemical Physics of Solids,
D-01187 Dresden, Germany}

\author{T.~Gruner}
\affiliation{Max Planck Institute for Chemical Physics of Solids,
D-01187 Dresden, Germany}

\author{C.~Krellner}
\affiliation{Max Planck Institute for Chemical Physics of Solids,
D-01187 Dresden, Germany}

\author{C.~Klingner}
\affiliation{Max Planck Institute for Chemical Physics of Solids,
D-01187 Dresden, Germany}

\author{C.~Geibel}
\affiliation{Max Planck Institute for Chemical Physics of Solids,
D-01187 Dresden, Germany}

\author{A.~Loidl}
\affiliation{Experimental Physics V, Center for Electronic Correlations and Magnetism,
University of Augsburg, D-86135 Augsburg, Germany}

\author{F.~Steglich}
\affiliation{Max Planck Institute for Chemical Physics of Solids,
D-01187 Dresden, Germany}

\date{\today}

\begin{abstract}
We investigate the electron-spin resonance (ESR) phenomenon in the heavy-fermion metal YbRh$_2$Si$_2$ by applying hydrostatic pressure up to 3~GPa and by inducing the internal pressure on the Yb site chemically in Yb(Rh$_{1-x}$Co$_x$)$_2$Si$_2$ samples. We found that the increase in pressure, reducing the hybridization between $4f$ and conduction electrons, leads to a remarkable change in the temperature dependence of the $g$ factor and broadens the ESR line. We relate the differences between the effect of internal and external pressure on the low temperature ESR parameters to the disorder induced by Co doping. The pressure effects on the Yb$^{3+}$ related resonance in YbRh$_{2}$Si$_{2}$ again manifest its local character when compared with the resonance of diluted Gd in heavy-fermion metals. 
\end{abstract}

\pacs{76.30.-v, 71.27.+a}

\maketitle

\section{Introduction}
The Kondo effect is now recognized to be of fundamental importance in a wide class of correlated electron systems. Experiments have demonstrated its significance not only in metals but also in nanoscale magnets, semiconductor quantum dots, carbon nanotubes and individual molecules.\cite{jarillo-herrero05a} The discovery of a well defined electron-spin resonance (ESR) signal in a series of Kondo-lattice systems below the Kondo temperature $T_{K}$ where heavy-fermion behavior begins to develop gives a unique opportunity to study the evolution of the Kondo state and heavy-fermion formation directly from the ESR measurables.\cite{sichelschmidt03a,krellner08a,schaufuss09a} This is a strong advantage in contrast to earlier ESR experiments in Kondo-lattices, which needed Gd doping as an ESR probe.\cite{krug98a} The origin and the unexpectedly small width of the absorption line are currently under intense theoretical investigations\cite{abrahams08a,zvyagin09b,schlottmann09a,huber09a,kochelaev09a,wolfle09a} which take into account that the ESR in Kondo-lattice systems is associated with the presence of ferromagnetic correlations\cite{krellner08a}. These theories follow approaches within a Fermi liquid \cite{abrahams08a} and non-Fermi liquid\cite{zvyagin09b,wolfle09a} description of itinerant heavy electrons, utilize a Kondo-lattice model  \cite{schlottmann09a}, focus on the ESR signatures of the strongly anisotropic Kondo-ion interactions within a molecular field description\cite{huber09a} or give a microscopic analysis of the spin dynamics of the Kondo-ions.\cite{kochelaev09a}\\
\indent
A particular interest was drawn to the ESR signal of the Kondo-lattice compound YbRh$_2$Si$_2$ \cite{sichelschmidt03a} which is located very close to a quantum critical point corresponding to the disappearance of weak antiferromagnetic (AFM) order. \cite{gegenwart08a} The signal shows pronounced properties of a localized Yb$^{3+}$ spin state but at the same time heavy-electron properties of the signal could be demonstrated by the signal dependence on the magnetic field.\cite{schaufuss09a} Conduction electron (CE) spins and Yb$^{3+}$ spins are strongly coupled and, hence, the ESR parameters were found to reflect changes in the hybridization between $4f$- and CE states when changing the unit-cell volume by substituting the Yb or Si sites with La or Ge, respectively.\cite{sichelschmidt05a, wykhoff07a} However, these data could not provide conclusive results about the relation of the resonance linewidth to the hybridization strength, the Kondo interaction, and disorder scattering, because in case of Ge doping only one concentration was available and La doping reduces the concentration of Yb$^{3+}$ spins. Interestingly, the latter influenced the linewidth in a way very reminiscent to a so-called ''bottleneck-relaxation'' mechanism  \cite{wykhoff07a} which was indicated also by the ESR of Lu doped YbRh$_{2}$Si$_{2}$ \cite{duque09a} and which was discussed extensively for the spin dynamics of \textit{diluted} magnetic moments in metallic hosts.\cite{barnes81a} A recent microscopic description of the spin dynamics in YbRh$_{2}$Si$_{2}$ reflects a bottleneck relaxation in a dense Kondo system with strong anisotropic exchange interactions: a collective spin motion of the Kondo ions with conduction electrons was shown to explain the narrow linewidth by virtue of the Kondo effect.\cite{kochelaev09a}\\
\indent
Here we report a systematic investigation of the ESR in YbRh$_{2}$Si$_{2}$ by a consecutive change in the hybridization strength by decreasing the unit-cell volume either by applying external hydrostatic pressure or by introducing chemical pressure by doping the Rh site with the isoelectronic but smaller Co ions. The latter allows to discriminate pure pressure effects from the effects of chemical modifications and disorder. As expected, pressure as well as Co doping stabilizes the antiferromagnetism in YbRh$_2$Si$_2$.\cite{friedemann09a}\\
\indent
Reports of ESR under pressure in metallic compounds are rare owing to the difficulty to achieve a high enough sensitivity. Up to now, there are only two studies available: the first one -- of the metal-insulator transition in Yb:Eu by Continentino {\it et al.} \cite{continentino95a}, the second one -- of the spin relaxation times in CeAl$_3$:Gd by Schlott {\it et al.} \cite{schlott90a}. It was shown that the external pressure strongly affects the hybridization strength, what leads to the opening of a conduction-band gap in ytterbium at 1.3~GPa and to an increase in the characteristic spin-fluctuation temperature $T_0$ in CeAl$_3$. In contrast to Ce systems, where the application of pressure strengthens the 4$f$-conduction electron hybridization and eventually generates valence fluctuations, the 4$f$ states in an Yb-based heavy-fermion system become more localized under pressure.\cite{goltsev05a} Such a picture successfully describes a \textit{decrease} in $T_0$ from 25~K to 8~K in YbRh$_{2}$Si$_{2}$ with pressures up to 1.7~GPa.\cite{mederle02a, tokiwa05a}

\section{Experimental Setup}
We measured the absorbed power $P$ of a transversal magnetic microwave  field (X-band, $\nu \approx 9.4$~GHz) as a function of an external, static magnetic field $H$. To improve the signal-to-noise ratio, a lock-in technique was used by modulating the static field, which yielded the derivative of the resonance signal $dP/dH$. All ESR spectra were obtained from platelet-shaped single crystals with $H\perp c$-axis of the tetragonal crystal structure. The samples were prepared and thoroughly characterized as described elsewhere.\cite{trovarelli00a,krellner09b,klingner09b}\\
\indent
High-pressure X-band ESR measurements were performed by means of a home-built hydraulic setup, which allows investigations in the pressure range up to 3~GPa.\cite{preusse87a} The sample was placed in a Cu-Be gasket filled with a methanol-ethanol mixture. The gasket was compressed by two opposed Al$_2$O$_3$ anvils, one of which served simultaneously as a dielectric cylindrical cavity, operated at a TE$_{111}$ mode. The use of a continuous-flow $^4$He cryostat allowed to perform measurements in the temperature range 2.5--300~K. The pressure was determined from the jump in the $ac$-resistivity at the superconducting transition of small pieces of lead which were put into the gasket together with the sample. The temperature was controlled with a calibrated carbon resistor. 
The X-band measurements of the Co doped single crystals were performed at ambient pressure with a standard Bruker Elexsys Spectrometer equipped with a $^{4}$He continuous gas-flow cryostat (Oxford instruments). 
%

\section{Results}
\label{results}
\begin{figure}[ht]
\centering
\includegraphics[width=0.8\linewidth]{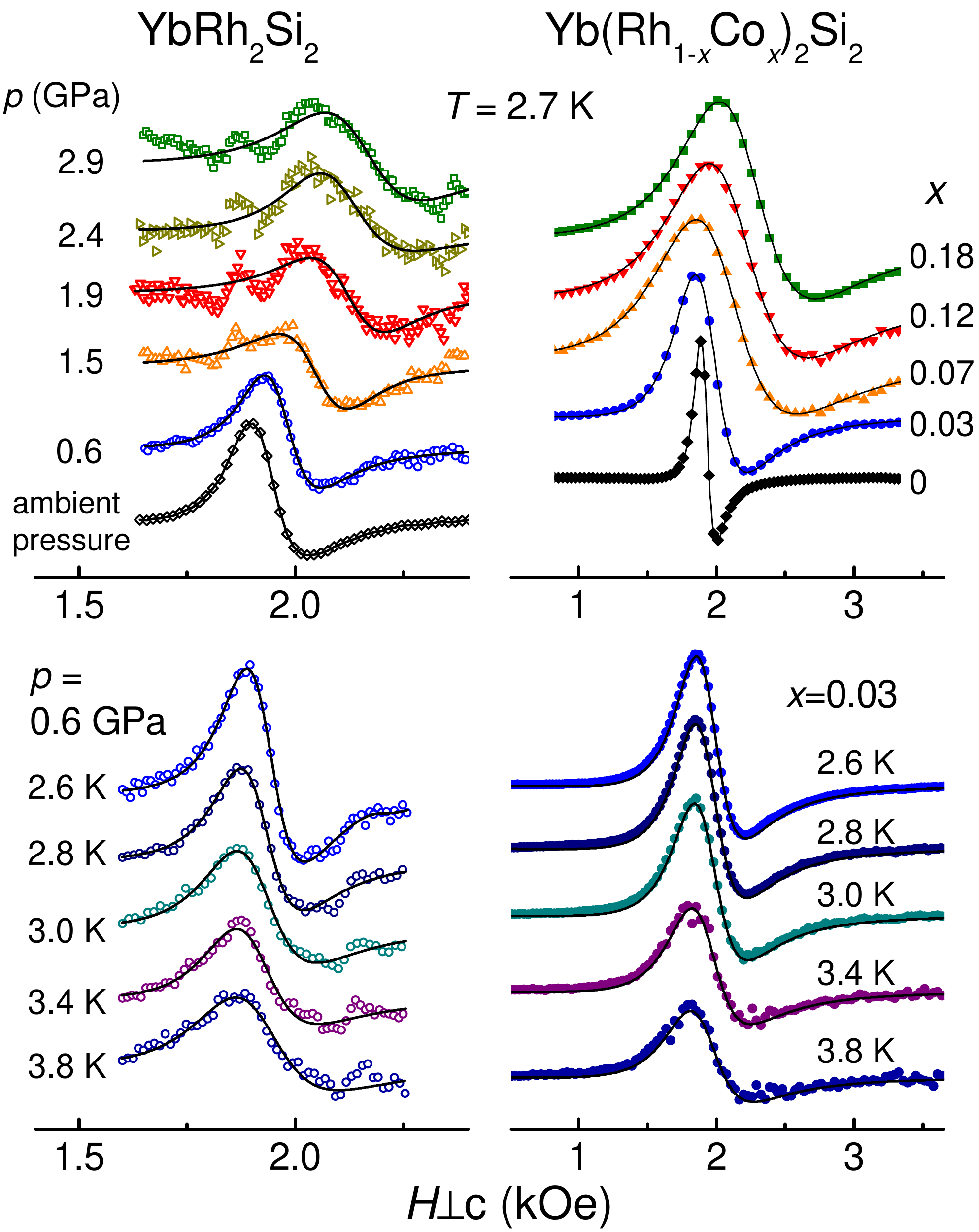}
\caption{(color online) ESR spectra $dP/dH$ (arbitrary units) of Yb(Rh$_{1-x}$Co$_x$)$_2$Si$_2$ at 9.4~GHz and $H\perp c$ with fitted Lorentzian shapes (solid lines). Amplitudes are scaled for best illustration. Top frames: $T = 2.7$~K at various hydrostatic pressures $p$ $(x=0)$ and Co-contents $x$ $(p=0)$. A weak signal near 1.8~kOe showing up for $p\ge 1.5$~GPa is due to the background of the pressure cell. Bottom frames: Temperature dependence for $p=0.6$~GPa $(x=0)$ and $x=0.03$ $(p=0)$.}
\label{Lines}
\end{figure}

In accordance with previous experiments \cite{sichelschmidt03a} all recorded ESR spectra consist of a single resonance line (see Fig.\ref{Lines}, symbols). We can satisfactorily describe the line by a Lorentzian line shape $dP/dH$ (Refs. \onlinecite{wykhoff07b} and \onlinecite{zakharov03a}) (frequently called ``Dysonian'', solid lines in Fig.~\ref{Lines}) which appears asymmetric due to the high sample
\begin{figure}[tb]
\centering
\includegraphics[width=0.8\linewidth]{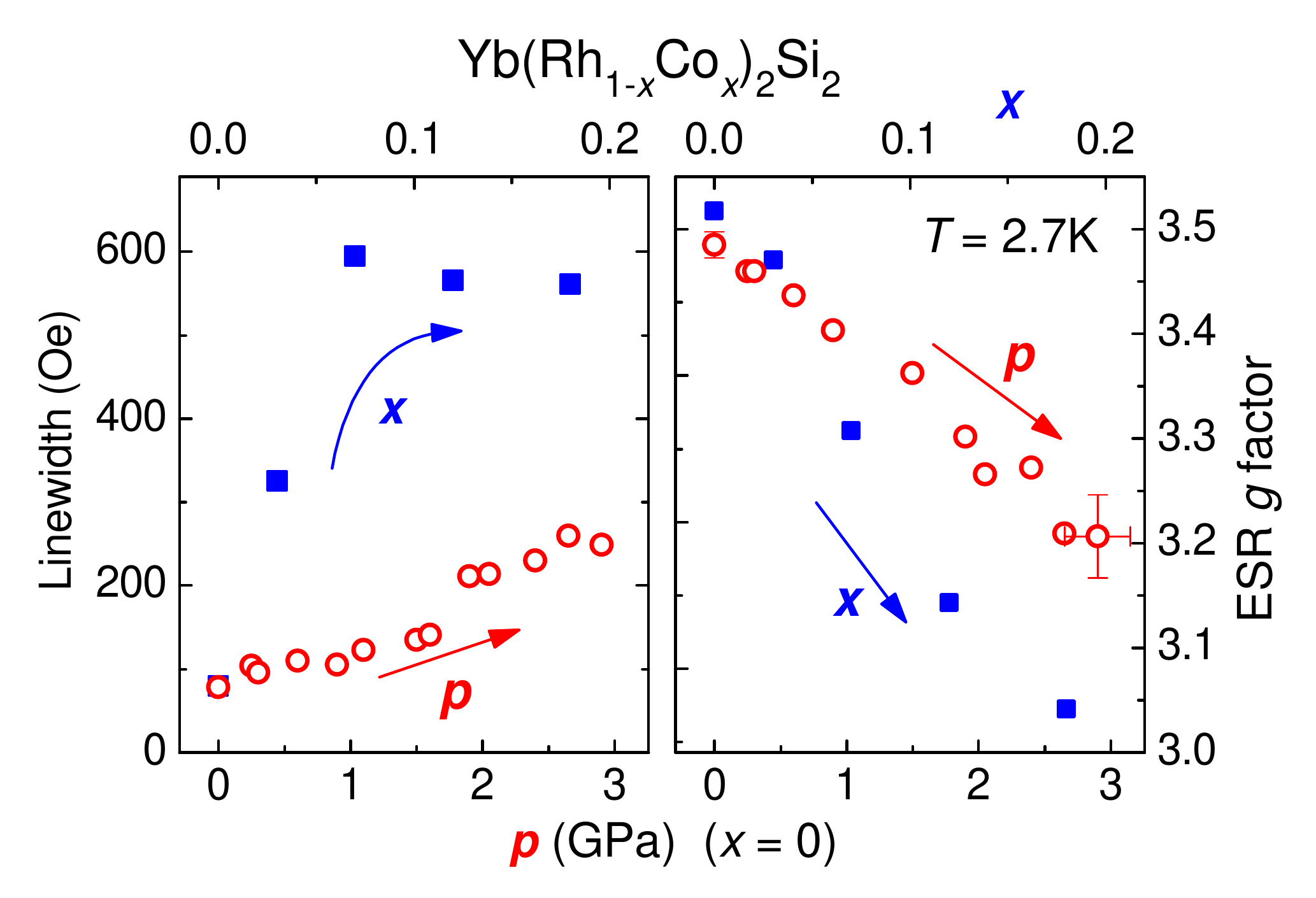}
\caption{ (color online)
Linewidth and effective $g$ factor of Yb(Rh$_{1-x}$Co$_x$)$_2$Si$_2$ at $T = 2.7$~K and $H\perp c$. Open circles indicate the effect of hydrostatic pressure $p$ for $x=0$ (lower axis), closed squares denote chemical pressure by a Co doping concentration $x$ at ambient hydrostatic pressure (upper axis). Error bars are given assuming the uncertainty in the resonance field as 10\% of the linewidth.
}
\label{PressDep}
\end{figure}
\begin{figure}[t]
\centering
\includegraphics[width=0.9\linewidth]{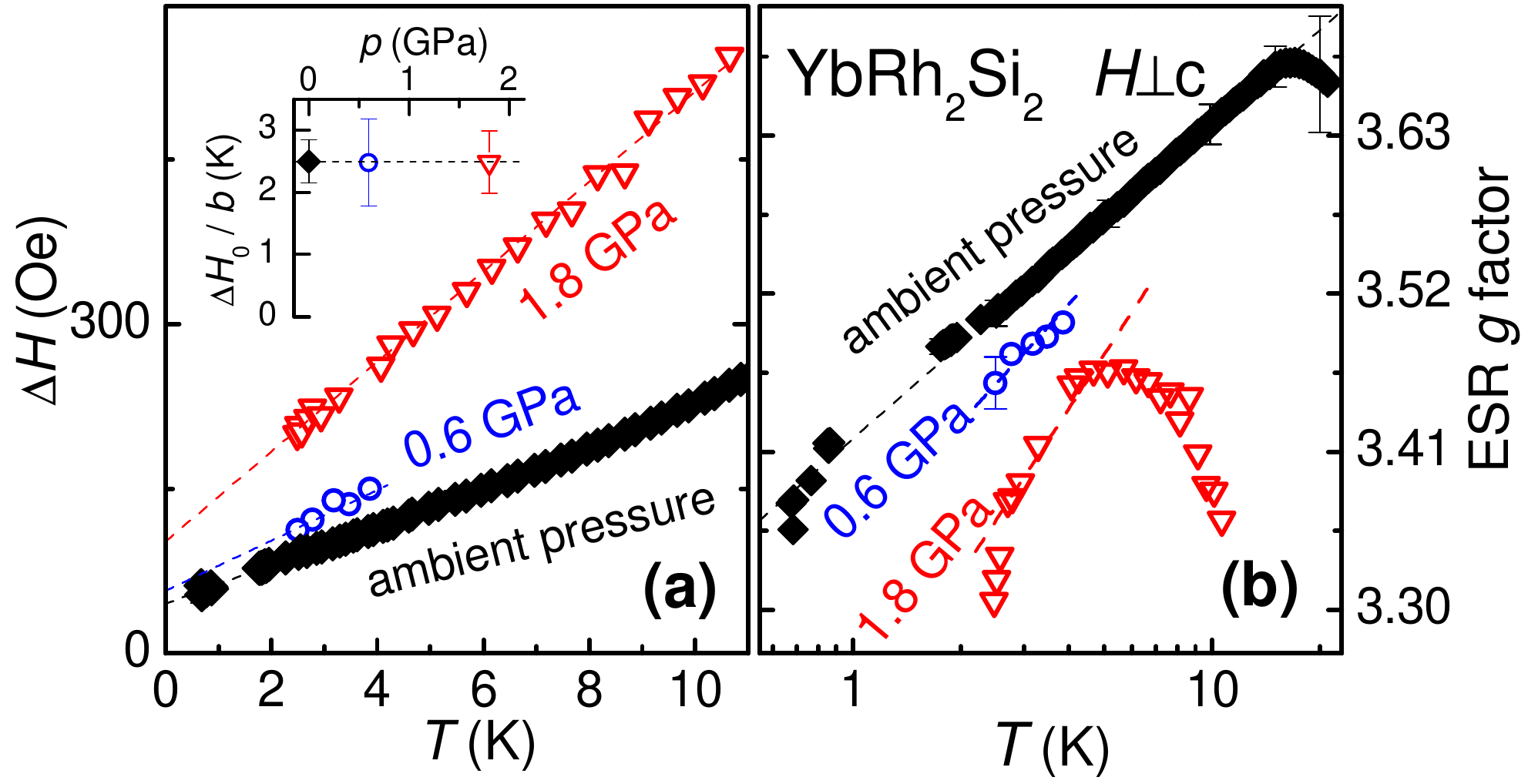}
\caption{(color online) Temperature dependence at various hydrostatic pressures and field $H\perp c$ of (a) the linewidth $\Delta H$ and (b) the effective ESR $g$ factor.
Dashed lines in (a) display $T$-linear behavior of $\Delta H$ with slopes $b$ and zero temperature value $\Delta H_0$ with $\Delta H_0/b\approx 2.5$~K as displayed in the inset. Dashed lines in (b) display $\log T$ dependencies of $g$ (lines are guides to the eyes).} \label{p_dHgTempDep}
\end{figure}
conductivity giving rise to the skin effect which admixes dispersion to the absorption spectra. The skin depth was larger than $1\mu\rm m$ but, for all samples, much smaller than the sample thickness. We kept the ratio of dispersion to absorption constant, which is reasonable regarding the fact that the skin depth remains by far smaller than the sample dimensions in the temperature range under consideration. The line fits provided the parameters resonance field  $H_{\rm res}$ and half-linewidth at half maximum, $\Delta H$.
As displayed in Fig.~\ref{Lines} for $T = 2.7$~K, both increasing pressure and Co concentration lead to a line broadening and a shift of $H_{\rm res}$ towards higher fields. External hydrostatic pressure qualitatively affects the ESR parameters in the same way as chemically induced pressure. For a pressure up to at least 0.6~GPa and a chemical pressure with a Co doping up to $x=0.03$ the corresponding ESR spectra are equivalent and can be described by the same $g$ factors, see Fig. \ref{PressDep}. In our pressure setup the ESR $g$ factor, as calculated from the resonance condition $h\nu=g\mu_BH_{\rm res}$ ($\mu_B$: Bohr magneton), amounts to $g_{\perp}=3.485(6)$ at ambient pressure and $T=2.7$~K. This agrees well with the previously published value \cite{sichelschmidt03a} in view of the different, higher quality batch of the crystal used here. 
 
\begin{figure}[b]
\centering
\includegraphics[width=0.9\linewidth]{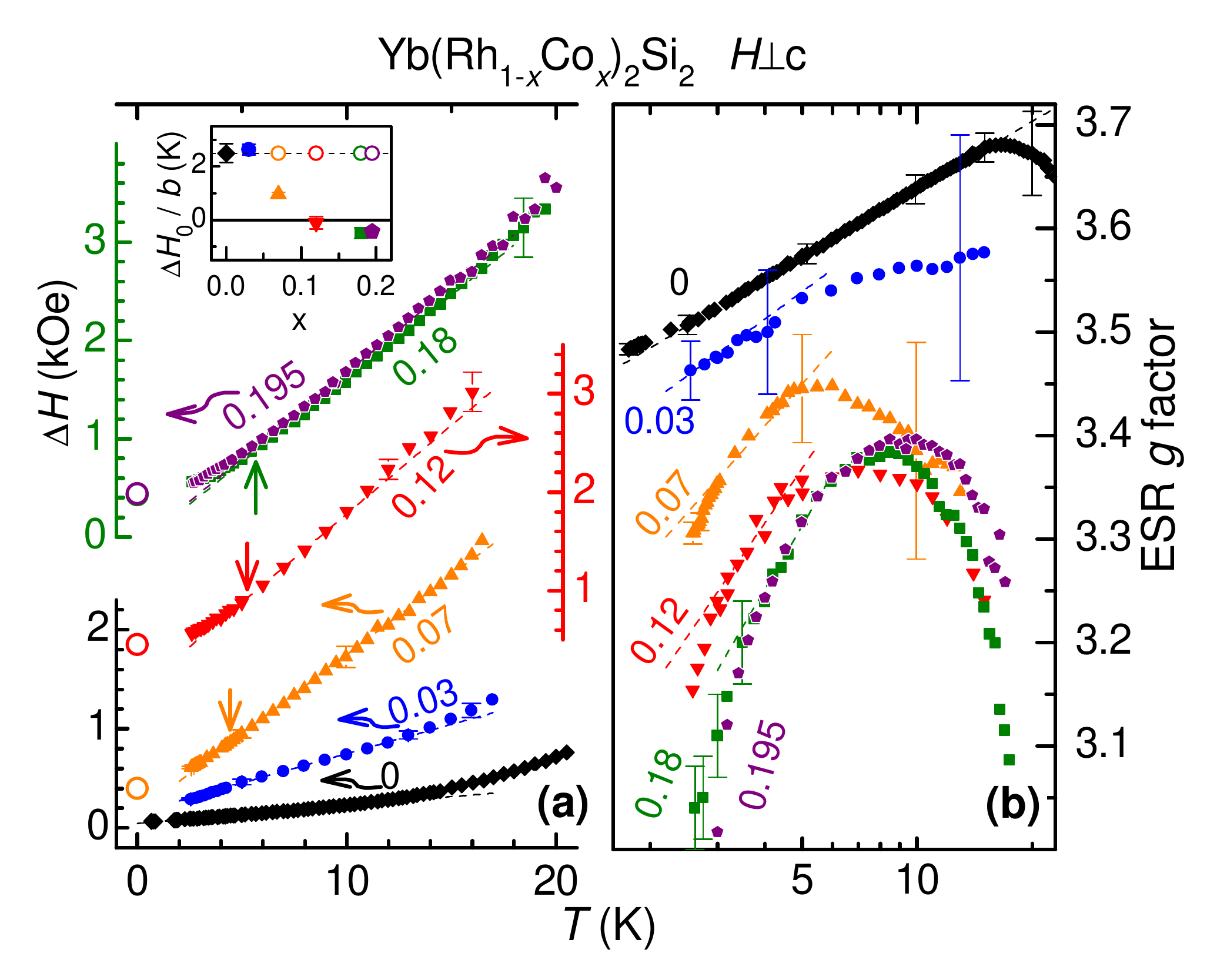}
\caption{(color online) Temperature dependence at various chemical pressures induced by Co content $x$ and field $H\perp c$ of (a) the linewidth $\Delta H$ and (b) the effective ESR $g$ factor. Dashed lines display (a) $T$-linear behavior of $\Delta H$ with slopes $b$ (deviation indicated by arrows) and (b) $\log T$ dependencies of $g$ (guide to the eyes). Inset displays the ratio of zero temperature linewidth $\Delta H_0$ and $b$ where $\Delta H_0$ arises from a linear linewidth extrapolation towards $T=0$ (solid symbols). The dashed line in the inset indicates $\Delta H_0/b\approx 2.5$~K (compare Fig. \ref{p_dHgTempDep}) which is used to determine the zero-temperature extrapolation of $\Delta H$ for $x\geq0.07$ denoted by open circles. } \label{Co_dHgTempDep}
\end{figure}

The dependence of the ESR parameters on hydrostatic pressure $p$ and chemical pressure by Co doping $x$ is displayed in Fig.~\ref{PressDep} for $T = 2.7$~K. 
The $x$- and $p$-values are related according the measured lattice parameters and the bulk modulus of YbRh$_{2}$Si$_{2}$.\cite{plessel03a} The relation could be confirmed by a perfect agreement of the transition temperatures of antiferromagnetic order of the Co-doped and pressurized samples.\cite{friedemann09a}
As shown with the open symbols, increasing hydrostatic pressure leads to an increase in the linewidth from 80~Oe at ambient pressure to 250~Oe at $p= 2.9$~GPa.  At the same time the $g$ factor shows an almost linear shift down to smaller values of 3.21(4) at $p = 2.9$~GPa. 
The $g$ factor in Yb(Rh$_{1-x}$Co$_x$)$_2$Si$_2$ coincides with the one in the parent compound at low Co concentrations $x < 0.07$, but deviates to lower values for $x > 0.07$. At $x=0.12$ it is shifted by $\approx-0.15$ from the value at the corresponding pressure $p=1.8$~GPa, independently on temperature within experimental error (compare Figs. \ref{p_dHgTempDep}b and \ref{Co_dHgTempDep}b). This indicates that the ionic $g$ value is reduced by Co doping and, concomitantly, the crystalline electric field should depend on Co doping as well.
The typical linewidths in the Co-doped compounds are considerably larger than the linewidths at corresponding pressures and saturate to a value of about 600~Oe for $x \ge 0.07$. In this respect one should note the linear relationship between the residual linewidth and the electrical resistivity which was found for La-doped YbRh$_{2}$Si$_{2}$ \cite{wykhoff07a} and which we discuss for the Co-doped YbRh$_{2}$Si$_{2}$ in Sec. \ref{disc}.   
\begin{figure}
\centering
\includegraphics[width=0.9\linewidth]{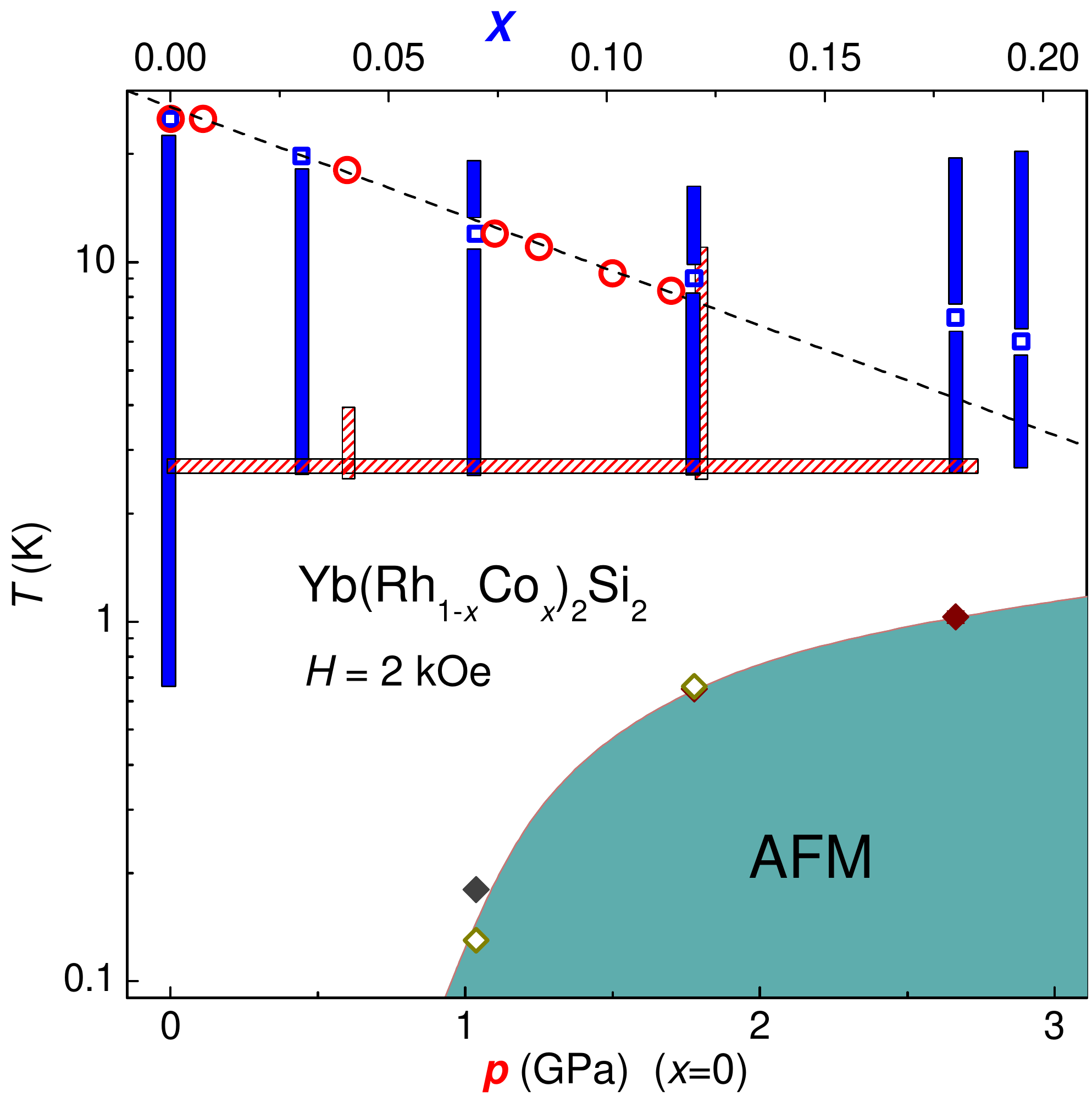}
\caption{(color online) Temperature-pressure phase diagram at $H=2$~kOe (ESR X-band resonance field at $g=3.4$) showing the performed ESR measurements for various Co contents $x$ (solid bars) and hydrostatic pressures $p$ (shaded bars). The $x$- and $p$-values are linearly related according to the measured lattice parameters and the bulk modulus of YbRh$_{2}$Si$_{2}$.\cite{friedemann09a} The dark shaded region indicates long-range AFM order (data from Refs. \onlinecite{mederle02a,krellner09b,klingner09a}). The Kondo temperature $T_{K}$ is shown by the open circles (under pressure) and open squares (with Co doping) (data from Refs. \onlinecite{tokiwa05a,krellner09b,klingner09b}). The dashed line shows a Kondo-type temperature dependence $\exp(-0.7\,{\rm GPa}^{-1}\cdot p)$.}   
\label{phdia}
\end{figure}
%
The temperature dependences of the ESR parameters were investigated at hydrostatic pressures of 0.6~GPa and 1.8~GPa and for Co concentrations up to $x=0.195$ (see illustration in Fig. \ref{phdia}). Figs. \ref{p_dHgTempDep}a and \ref{Co_dHgTempDep}a show the results for the ESR linewidth. The dashed lines indicate a linear law $\Delta H = \Delta H_0 + bT$ in a large $T$-region, reminding to a Korringa law of local moment relaxation towards conduction electrons \cite{barnes81a} with parameters listed in Tab. \ref{table1}. 
\begin{table}[htdp]
\begin{center}
\begin{tabular}{c|c|c|c|c}
$p$ (GPa) & $x$ & $\Delta H_0$ (kOe) & $b$ (Oe/K) & $\Delta H_0/b$ (K) \\ \hline
ambient & 0 & 0.045 & 18 & $2.5\pm0.35$\\
0.6 & 0 & 0.057 & 23 & $2.5\pm0.7$\\
1.8 & 0 & 0.102 & 41& $2.5\pm0.5$ \\ \hline
ambient & 0.03 & 0.156 & 59 & $2.6\pm0.2$\\
ambient & 0.07 & 0.150 & 160 & $0.95\pm0.1$\\
ambient & 0.12 & -0.02 & 180& $-0.1\pm0.2$ \\
ambient & 0.18 & -0.09 & 170 & $-0.5\pm0.2$\\
ambient & 0.195 & -0.075 & 175 & $-0.4\pm0.2$
\end{tabular}
\end{center}
\caption{Linewidth parameters of $\Delta H = \Delta H_0 + bT$ obtained from linear fits as shown by the dashed lines in Figs. \ref{p_dHgTempDep}a and \ref{Co_dHgTempDep}a.}
\label{table1}
\end{table}
Notable differences from the Korringa behavior occur above 12~K for $x=0$, indicating the influence of the first excited crystal-field level.\cite{sichelschmidt03a} Interestingly, the linewidth data can be characterized by a ratio $\Delta H_0/b$ which indicates an interesting universal feature of the ESR relaxation as depicted in the insets of Figs. \ref{p_dHgTempDep}a and \ref{Co_dHgTempDep}a: $\Delta H_0/b \approx 2.5$~K remains almost unaffected by pressure and by Co doping up to $x=0.03$. Such behavior was previously already identified as a linewidth scaling of various YbRh$_{2}$Si$_{2}$ batches (which differ in their residual resistivities) and of La-doped YbRh$_2$Si$_2$ pointing to a common relaxation mechanism to which both $\Delta H_0$ and $b$ can be ascribed.\cite{wykhoff07a}
%
For the ESR linewidth data of Co concentrations $x\geq0.07$ two characteristic differences to the corresponding pressure data are observable: Firstly, a deviation from linearity occurs in the low $T$-region which is marked by arrows in Fig. \ref{Co_dHgTempDep}. This deviation appears as soon as antiferromagnetic order at much lower temperatures has been established by Co doping, see phase diagram in Fig. \ref{phdia}.
%
Secondly, the ratio $\Delta H_0/b$ strongly deviates from $2.5\,{K}$ and even \textit{negative} values of $\Delta H_0$ appear for $x\geq0.12$ (see Tab. \ref{table1}).  
If one assumes the above mentioned universal value $\Delta H_0/b \approx 2.5$~K to hold also for Yb(Rh$_{1-x}$Co$_x$)$_2$Si$_2$ with $x\geq 0.07$, one obtains values $\Delta H_0'$ which reasonably agree with the linewidth data at the lowest accessible temperatures (open circles in Fig. \ref{Co_dHgTempDep}a).

%
The temperature dependence of the $g$ factor under hydrostatic as well as chemical pressure is illustrated in Figs.~\ref{p_dHgTempDep}b and \ref{Co_dHgTempDep}b. In the low-temperature region the dashed lines emphasize a logarithmic temperature dependence which could well describe the results for YbRh$_2$Si$_2$ at ambient pressure as was previously reported.\cite{sichelschmidt03a}  
In the presence of antiferromagnetic order for pressures larger than 1~GPa (see Fig. \ref{phdia}) $g(T)$ deviates from a logarithmic decrease to a much stronger down-turn, see the 1.8 GPa data in Fig. \ref{p_dHgTempDep}b. In this respect two peculiarities are worth to note. For the datasets $x=0.12$ and the chemical pressure equivalent $p=1.8$~GPa the deviation appears at the same temperature of about 2.5~K, see Figs.~\ref{p_dHgTempDep}b and \ref{Co_dHgTempDep}b. Furthermore, in case of the Co doped samples these deviations appear at about the same temperatures where the linewidth departs from the linear temperature dependence, see arrows in Fig. \ref{Co_dHgTempDep}a.
%

\section{Discussion}
\label{disc}
%
The phase diagram shown in Figure \ref{phdia} displays the parameter ranges of the ESR experiments under pressure (shaded bars) and Co doping (solid bars) together with magnetic and electronic properties. Applying pressure (either external hydrostatically or chemically by Co doping) suppresses the fluctuations of the Yb-valence and leads to a decrease in the Kondo temperature $T_{K}$.\cite{goltsev05a} This effect is shown by the open circles and open squares which correspond to the characteristic spin-fluctuation temperature $T_0$ in the 4$f$-increment of the low-temperature electronic specific heat $\Delta C/T\propto -\ln(T/T_0)$. At ambient pressure and zero field $T_0$ was shown to match with the single-ion Kondo temperature (obtained by evaluating the 4$f$ entropy)\cite{gegenwart06a} and, hence, the temperature dependence of $T_0$ is consistent with an exponential law shown by the dashed line: $T_{K}\propto\exp[-1/JN_c(E_{\rm F})]$ where $J$ is the exchange integral between 4$f$ electrons and conduction electrons and $N_c(E_{\rm F})$ is the conduction electron density-of-states at the Fermi surface, and assuming $x,p\propto 1/JN_c(E_{\rm F})$.
Figure \ref{Lines} demonstrates the qualitatively corresponding effect of pressure and Co doping on the ESR results which confirms that pressure effects are amenable with our ESR setup. 
The pressure effect on the temperature dependence of the linewidth can nicely be compared with the pressure effect on the ESR relaxation of diluted Gd$^{3+}$ in Ce-Kondo-lattice systems, in CeAl$_3$ for instance.\cite{schlott90a} There, both the magnetic susceptibility $\chi_{\rm Ce}$ and the fluctuation time $\tau\propto 1/T_{\rm K}$ of the Ce-4$f$ moments decrease by application of pressure. This effect could be detected in the low temperature relaxation of the Gd$^{3+}$ resonance: with increasing pressure the linewidth shows a decrease in the linear temperature slope $b\propto \chi_{\rm Ce}/T_{\rm K}$ and, simultaneously, a decrease in $\Delta H_0$. 
 A corresponding pressure effect, transferred to Yb-systems,\cite{goltsev05a} is reflected by the linewidth behavior of YbRh$_{2}$Si$_{2}$. In agreement with the pressure-induced decrease in $T_{\rm K}$ (dashed line in Fig. \ref{phdia}) the slope $b$ of the linewidth is increased by pressure and, also, a simultaneous increase in $\Delta H_0$ is observed (see Figs. \ref{p_dHgTempDep}a and \ref{Co_dHgTempDep}a). This equivalency points out that the ESR in YbRh$_{2}$Si$_{2}$ shows features of a local Yb$^{3+}$ spin relaxation being influenced by the spin dynamics of the surrounding Yb$^{3+}$ ions. 
A theoretical basis for understanding such signatures of locality in the linewidth-pressure dependences may be supplied by a recently developed model of the collective spin motion of 4$f$-  and conduction-electron spins in a Kondo lattice with strongly anisotropic Kondo interactions.\cite{kochelaev09a} It reveals for the bottleneck regime a collective spin mode with a narrow linewidth showing the observed temperature dependence. In this model, besides the strong reduction in the linewidth by virtue of the Kondo effect, the line experiences also a ``motional'' narrowing process due to the translational diffusion of quasilocalized $f$-electrons in the non-Fermi liquid state. If the RKKY interaction between nearest Yb ions is ferromagnetic, the diffusion process is supported and, hence, the linewidth depends also on the presence of short-range ferromagnetic fluctuations in YbRh$_{2}$Si$_{2}$. The application of pressure stabilizes antiferromagnetic correlations, leading to more localized $f$ electrons and reduces the efficiency of the motional narrowing process.
Within the phenomenology of a heavy-quasiparticle spin resonance the effect of ferromagnetic fluctuations on the quasiparticle scattering and spin-lattice relaxation was also found to significantly narrow the linewidth.\cite{wolfle09a} 
Furthermore, this framework stresses the role of the lattice coherence of the quasiparticles for effectively preventing a strong local relaxation of the Kondo spin. Therefore, regarding the different effect of Co doping and pressure for the absolute linewidth values one would expect a line-broadening once the lattice coherence is disturbed by Co doping which, indeed, is observed, see Fig. \ref{PressDep}.   
However, both approaches, describing the spin dynamics with a collective spin mode or within a quasiparticle picture, have not yet explicitly considered the pressure variation in the Kondo temperature in their expressions for the linewidth and $g$ factors.

%
Experimentally, a pressure-dependent Kondo scale is suggested by the observed dome-shape of the $g$ factor temperature dependence by application of pressure.
With increasing temperature $g(T)$ reveals a logarithmic-like increase in a limited region of temperature which is followed by a substantial decrease. This happens at a temperature which is largest at ambient pressure and $x=0$ suggesting a relation of the dome position and the Kondo temperature. The theoretical considerations for a heavy-quasiparticle spin resonance indeed qualitatively predict such a dome in $g(T)$ in the non-Fermi liquid regime which is determined by the temperature dependence of the quasiparticle effective mass and a contribution from a small anisotropy in the spin-exchange.\cite{wolfle09a} From the low temperature behavior of the $g$ factor the collective spin-mode approach reveals a characteristic temperature for the ground Kramers doublet which is by two orders of magnitude smaller than the Kondo temperature.\cite{kochelaev09a}   

\begin{figure}[tb]
\centering
\includegraphics[width=0.9\linewidth]{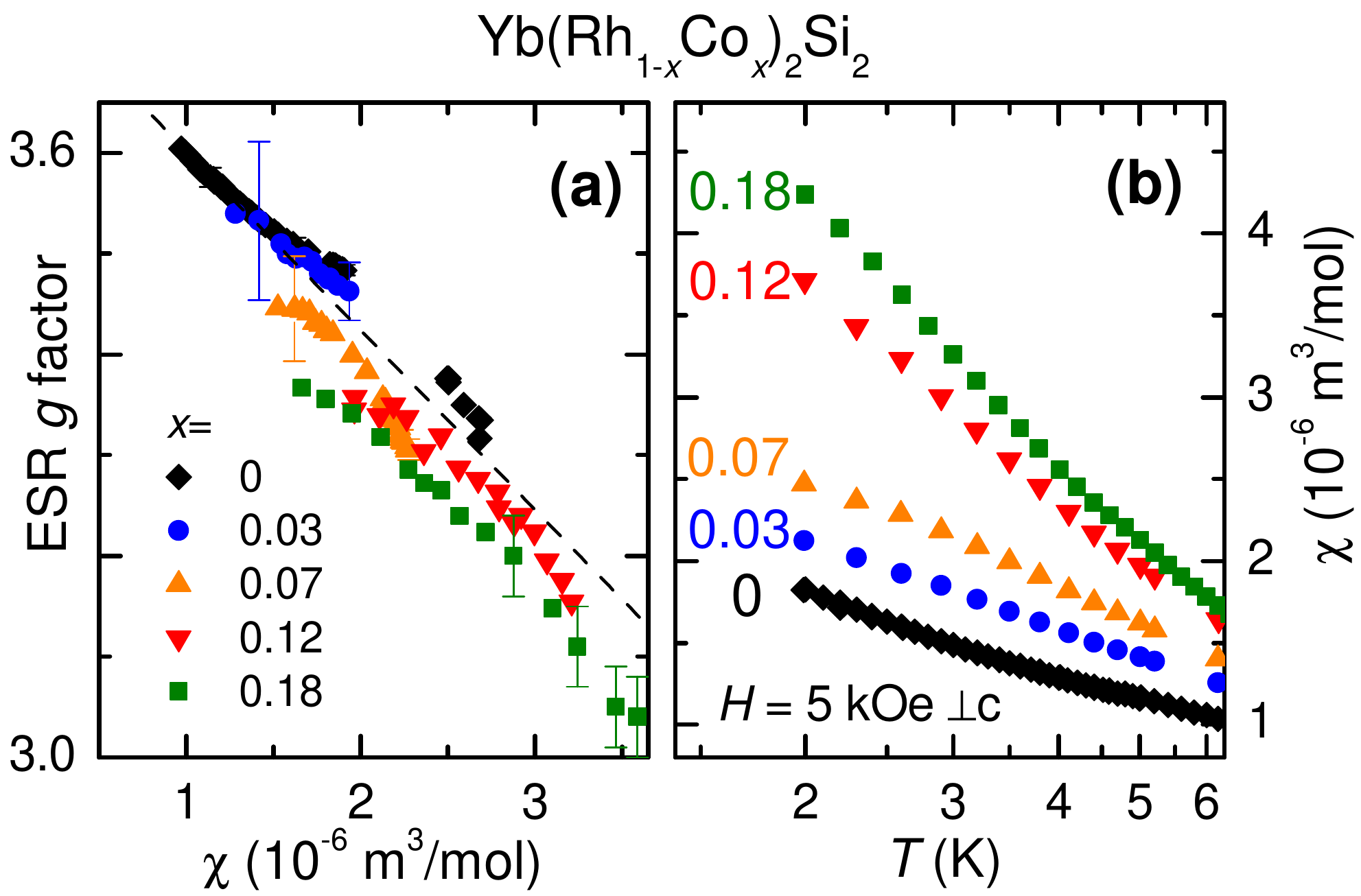}
\caption{(color online) (a) ESR $g$ values vs. magnetic susceptibility $\chi$ indicates a linear relationship (dashed line: fit as described in the text) for different chemical pressures induced by Co content $x$. (b)  $\chi (H=5$~kOe) vs. temperature (log scale) at different chemical pressures.} \label{gchi}
\end{figure}

For all pressures the observed ESR $g$ factors drop considerably below the insulator value $g_{0}$ expected for Yb$^{3+}$ in YbRh$_{2}$Si$_{2}$. For the previous results at ambient pressure this negative shift was related to a negative, \textit{antiferromagnetic} effective exchange coupling and a direct relation between the $g$ factor and the static magnetic susceptibility was found.\cite{sichelschmidt03a}
Figure \ref{gchi} shows that this is also the case for the Co-doped samples (appropriate susceptibility data under pressure were not available): the $g$ values for various Co contents are shifted to smaller values proportional to the susceptibility. 
Such behavior cannot be explained by the effect of demagnetization which leads to a much weaker influence of the $g$-factor ($\Delta g/ g < 0.5$\% for $\Delta T = 1$\,K) and which, moreover, would correspond to an opposite temperature dependence of the $g$-factor. Therefore, the temperature-dependent shift of the resonance could mostly be determined by the temperature dependence of internal antiferromagnetic exchange fields. A corresponding molecular-field approximation of the effective $g$ factor, $g_{\rm eff}=g_0[1+\lambda\chi(T)]$, is indicated by the dashed line in Figure \ref{gchi}a with $g_0=3.78$ \cite{kutuzov08a} and the molecular field parameter $\lambda=-3.3\;{\rm kOe}/\mu_{B}$ which is consistent with the $\lambda$-values estimated in Ref. \onlinecite{duque09a}. With increasing Co content from $x=0$ to $x=0.18$ one observes a parallel-like down-shift of the $g(\chi)$ curves which corresponds to a decrease in $g_0$ of approximately $3\%$ while a change in $\lambda$ could not be resolved within the given experimental accuracy. The decrease in $g_0$  could be caused by changing the crystalline electric field with Co doping as was mentioned above, regarding the differences between the pressure- and doping-induced $g$ shifts, see Fig. \ref{PressDep}. The close relationship between the $g$ factor and the magnetic susceptibility points out that the resonance reflects the local Yb$^{3+}$ magnetic properties directly. 

%
\begin{figure}[tb]
\centering
\includegraphics[width=0.9\linewidth]{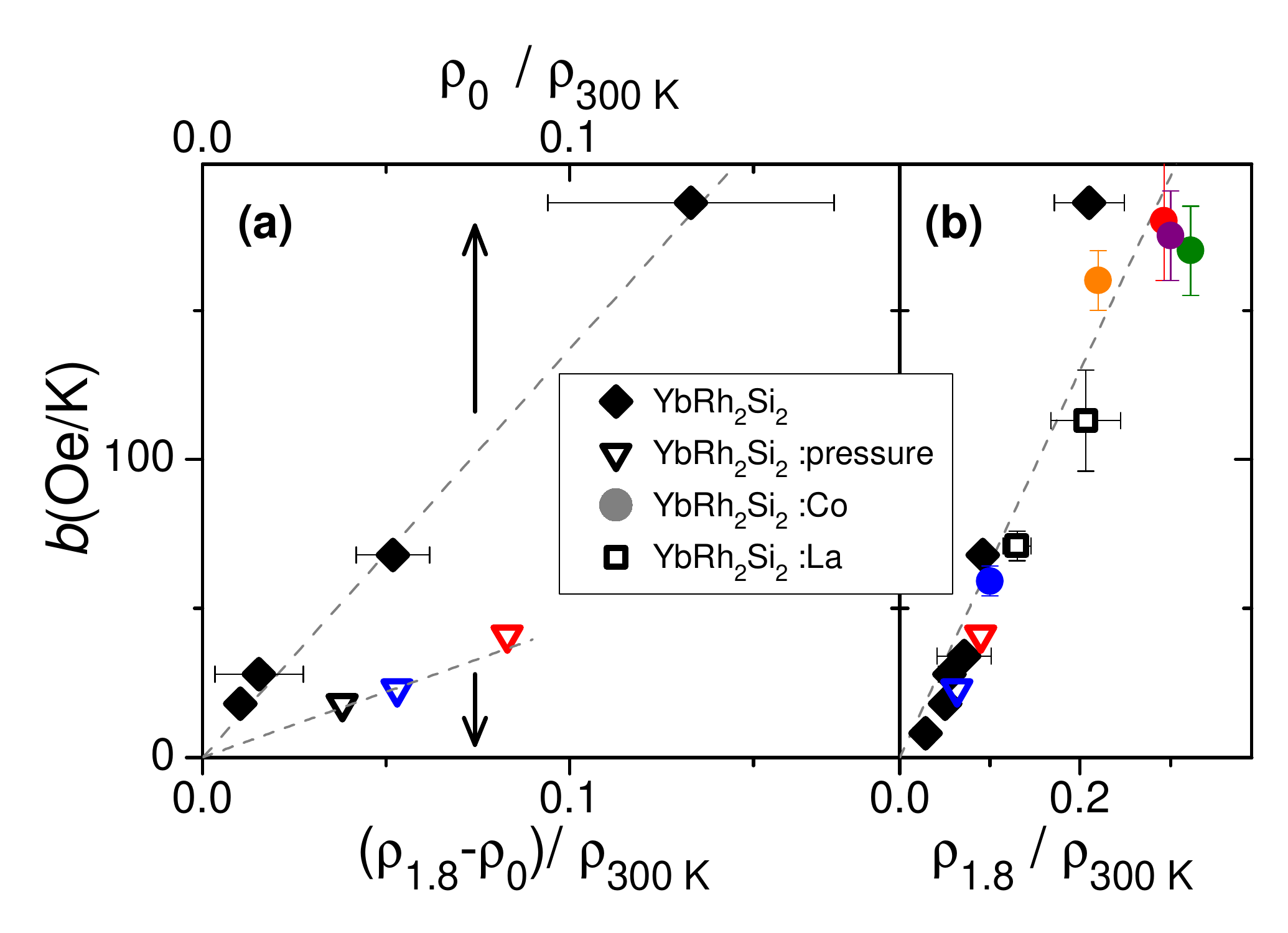}
\caption{(color online) Variation of the linear temperature linewidth slope $b$ with the ratios of electrical resistivity $\rho$ at temperatures 1.8~K, 300~K and 0.2~K ($\rho_{0}$). Closed diamonds correspond to YbRh$_{2}$Si$_{2}$ from different batches of indium flux grown crystals. Circles, triangles, and squares belong to samples with Co-doping, under pressure, and with La-doping \cite{wykhoff07a}, respectively. Dashed lines are guides to the eyes.} \label{dHRRR}
\end{figure}
As was shown in Sec. \ref{results}, the linewidth behavior under pressure is characterized by $\Delta H_0/b\approx 2.5$~K which provides evidence for a common relaxation mechanism involved in the residual linewidth $\Delta H_0$ and the slope of the linear temperature dependence $b$. 
In this respect it is interesting to note the relation of both quantities to the residual electrical resistivity indicating the relevance of charge transport scattering processes for the ESR relaxation in YbRh$_{2}$Si$_{2}$. This was previously shown for $\Delta H_0$ in Ref. \onlinecite{wykhoff07a} and is illustrated for $b$ in Figure \ref{dHRRR}.
%
In order to discriminate the scattering effects from disorder and Kondo interaction, Figure \ref{dHRRR}a compares $b$ as a function of the disorder-dominated $\rho_{0}=\rho(0.2\rm\,K)$ (upper axis) with the Kondo-dominated $\rho_{1.8\rm\,K}-\rho_{0}$ (lower axis). Whereas the effect of disorder variation should only show up among different YbRh$_{2}$Si$_{2}$ samples, the variation in the Kondo interaction should be dominant for the samples at various pressures. Indeed, as shown in Fig. \ref{dHRRR}a, this situation is evidenced for the $b$-values by their linear relation with the respective resistivity data. For the samples with La- and Co-doping both disorder scattering and Kondo interaction contribute to $b$ which then appears linearly related to $\rho_{1.8\rm\,K}/\rho_{300\rm\,K}$ as shown in Fig. \ref{dHRRR}b. In particular, with $\Delta H_0'\approx b\cdot 2.5$~K, this behavior is also observable for the Co-doped samples with $x\ge0.07$ and, thus, the extrapolation of the residual linewidth values $\Delta H_0'$ shown by the open circles in Fig. \ref{Co_dHgTempDep}a is also supported by $b\propto \rho_{1.8\rm\,K}/\rho_{300\rm\,K}$.

%
%
Nevertheless, the difference between the application of hydrostatic and chemical pressure for the electrical resistivity does not fully relate to the difference for the ESR parameters. This is most clearly indicated by the behavior of the linewidth towards low temperatures where the continuous linearity of the pressure data is in contrast to the presence of a kink in the Co data for $x\geq0.07$ (arrows in Fig. \ref{Co_dHgTempDep}a). Interestingly, $x\geq0.07$ also marks the lower bound for long-range antiferromagnetic order (in the presence of an X-band resonance field of $\approx 2$~kOe) as shown in the $p-T$ phase diagram of Fig. \ref{phdia} by the dark shaded area. Usually, when approaching magnetic ordering by lowering the temperature a slowing down of spin fluctuations results in a reduced narrowing process in the linewidth, i.e., in its increase. This provides a larger contribution from inhomogeneous broadening and is observed, for instance, for the resonance of Gd$^{3+}$ in Ce(Cu$_{1-x}$Ni$_{x}$)$_2$Ge$_2$ ($x<0.7$).\cite{krug98a} However, whether magnetic ordering is the dominant source for the arrow marked deviations of the $x\geq0.07$ data, is questionable regarding the lack of deviation from linearity in the pressure linewidth data. Also, the relation $\Delta H_0/b\approx 2.5$~K provides evidence that inhomogeneous broadening does not dominate $\Delta H_{0}$ in the regime where AFM order occurs at low temperatures. 
These considerations point out that investigating the spin dynamics in YbRh$_{2}$Si$_{2}$, as characterized by the ratio $\Delta H_0/b \approx 2.5$~K, is better achieved by the ESR under pressure than by the ESR with Co-doping where the spin dynamics is obscured by additional disorder-related effects.

\section{Conclusion}

We have studied the effect of hydrostatic pressure $p < 2.9$~GPa and Co doping $0\leq x \leq 0.195$ on the ESR in Yb(Rh$_{1-x}$Co$_x$)$_2$Si$_2$. Both, hydrostatic pressure and chemical pressure by Co doping lead to qualitatively the same effect, namely, an increase of the linewidth and a decrease of the $g$ factor under pressure while at the same time the Kondo temperature is decreased. 
The pressure dependence of the linear temperature slope of the linewidth is equivalent to the linewidth behavior of local Gd$^{3+}$ spins serving as diluted ESR probes in Ce-based heavy-fermion compounds.\cite{schlott90a} There, the change in the low-temperature slope under pressure is explained by the change in the Kondo temperature. Therefore the equivalency reveals two conclusions: (i) the ESR in YbRh$_{2}$Si$_{2}$ looks alike a resonance of local Yb$^{3+}$ spins in a metallic environment and (ii) the Kondo temperature is a relevant parameter to describe the linewidth. The present theoretical frameworks for the ESR in YbRh$_{2}$Si$_{2}$ \cite{kochelaev09a,wolfle09a} provide a reasonable basis to understand the ESR under pressure in terms of the Kondo effect and the presence of ferromagnetic correlations.  
The different effect of Co doping and pressure on the linewidth suggests that disorder induced by Co doping is more effective in destroying the lattice coherence, the latter being essential for observing narrow ESR lines in dense Kondo-lattice systems.\cite{wolfle09a}
The linewidth data for all investigated pressures and Co contents could be characterized by a universal ratio between the residual linewidth and the slope of the linear temperature dependence of the linewidth. By relating both quantities to the scattering processes of charge transport the evolution of the ESR data with pressure allow a further characterization of the influence of the Kondo interaction to the ESR of YbRh$_{2}$Si$_{2}$.

 
%

\indent This work was supported by the DFG within TRR 80 (Augsburg, Munich), by the Volkswagen Foundation (Grant No. I/82203 and I/84689), and the DFG Research Unit 960 ``Quantum Phase Transitions''. We acknowledge fruitful discussions with M. Brando.
%

\end{document}